\documentclass[12pt]{article}

\usepackage{bbm}
\usepackage{epsfig}
\usepackage{array}
\usepackage{float}
\usepackage{dsfont}
\usepackage{amstext}
\usepackage{rotating}
\usepackage{a4}
\usepackage{a4wide}
\usepackage{cite}

\def\ra{\rightarrow}
\def\be{\begin{equation}}
\def\ee{\end{equation}}
\def\gs{\mathrel{
   \rlap{\raise 0.511ex \hbox{$>$}}{\lower 0.511ex \hbox{$\sim$}}}}
\def\ls{\mathrel{
   \rlap{\raise 0.511ex \hbox{$<$}}{\lower 0.511ex \hbox{$\sim$}}}}

\newcommand{\ba}{\begin{array}{c}}
\newcommand{\baz}{\begin{array}{cc}}
\newcommand{\bad}{\begin{array}{ccc}}
\newcommand{\bav}{\begin{array}{cccc}}
\newcommand{\baf}{\begin{array}{ccccc}}
\newcommand{\bag}{\begin{array}{cccccc}}
\newcommand{\bea}{\begin{equation} \begin{array}{c}}
\newcommand{\eea}{ \end{array} \end{equation}}
\newcommand{\ea}{\end{array}}
\newcommand{\D}{\displaystyle}

\begin{document}

\title{
\hfill {\small arXiv: 0909.1219 [hep-ph]} 
\vskip 0.5cm
\Large \bf
Flavor Composition of UHE Neutrinos at 
Source and at Neutrino Telescopes} 
\author{
Sandhya Choubey$^a$\thanks{email: \tt sandhya@hri.res.in}~~,~~
Werner Rodejohann$^b$\thanks{email: 
\tt werner.rodejohann@mpi-hd.mpg.de} 
\\\\
{\normalsize \it$^a$Harish--Chandra Research Institute,}\\
{\normalsize \it Chhatnag Road, Jhunsi, 211019 Allahabad, India }\\ \\ 
{\normalsize \it$^b$Max--Planck--Institut f\"ur Kernphysik,}\\
{\normalsize \it  Postfach 103980, D--69029 Heidelberg, Germany} 
}
\date{ \today}
\maketitle
\thispagestyle{empty}
\vspace{-0.8cm}
\begin{abstract}
\noindent  
We parameterize the initial flux composition of high energy 
astrophysical neutrinos as $(\Phi_e^0 : \Phi_\mu^0 : \Phi_\tau^0) 
= (1 : n : 0)$, where $n$ characterizes the source. 
All usually assumed neutrino sources appear as limits of this simple
parametrization. We investigate 
how precise neutrino telescopes can pin down the value of $n$. 
We furthermore show that there is a neutrino mixing scenario in 
which the ratio of muon neutrinos to the other neutrinos takes a 
constant value {\it regardless of the initial flux composition}. This occurs when 
the muon neutrino survival probability takes its minimal allowed
value. The phenomenological consequences of this very predictive 
neutrino mixing scenario are given.

\end{abstract}

\newpage

\section{\label{sec:intro}Introduction}

Neutrino mass and mixing influences many phenomenological aspects of 
particle and astroparticle physics.  In particular, any 
observed flavor mix of neutrinos $\nu_e, \nu_\mu$ and $\nu_\tau$ is
usually not the original flavor mix, but rather a modified one. This 
is a consequence of the non-trivial structure of the 
leptonic mixing, or the Pontecorvo-Maki-Nakagawa-Saki (PMNS) mixing
matrix. In this letter we wish to focus on 
the interplay of neutrino mixing and the flavor composition 
of ultra high energy (UHE) astrophysical neutrinos, whose 
detection is the goal of neutrino telescopes such as 
IceCube \cite{ice}, currently under construction, or the 
planned KM3Net facility \cite{km3net}.  
Many recent works discussed aspects of this problem 
\cite{1,2,3,zz,4,xing,5,6,WR,7,lipari,8,9,10,PRW0,PRW1,CNR,new,NP}, 
an overview can be found in \cite{over}. 

The composition of the flux detectable in neutrino telescopes 
depends on two things -- (i) the initial flavor composition at the 
source, and (ii) the standard (and 
possibly non-standard) parameters of neutrino 
mixing. This means that there are two 
types of studies one may pursue with neutrino telescopes: 
\begin{itemize}
\item[(a)] Neutrino physics: the extraction of neutrino properties 
from measurements of flavor ratios; 
\item[(b)] Astrophysics: the identification of the initial 
flavor composition of the flux of UHE neutrinos and using that 
to probe the type of source and its properties.
\end{itemize}
A large plethora of literature exists 
for item (a), where authors have studied the potential of probing the 
neutrino mixing angles using the flavor ratios. 
In this paper we will concentrate on item (b) and 
expound the potential of UHE neutrino 
measurements to decipher the flavor composition of the UHE 
neutrinos at their source (see Ref.~\cite{new} for related 
recent analyses). This could lead to a better 
understanding of 
the properties of the astrophysical source responsible for the 
production of these neutrinos. As will be discussed in detail 
in Section 3, different astrophysical situations could give rise 
to a variety of flavor ratios of the neutrinos. Since the  
usually assumed high energy 
neutrino sources do not generate tau neutrinos in any appreciable  
amounts, we neglect their presence in the initial UHE neutrino flux 
and propose a simple 
one parameter parameterization of the initial flux composition 
of the neutrinos: 
\be \label{eq:main}
(\Phi_e^0 : \Phi_\mu^0 : \Phi_\tau^0) = (1 : n : 0) \, .
\ee
All discussed sources such as pion, muon-damped, charm or
neutron beams are simple limits of this parameterization. 
With such a parameterization the experimental determination of 
the initial flux composition is simply an extraction of the 
parameter $n$ from the observed flux composition on Earth. 
We will define two kinds of measurable flux ratios 
and calculate the predicted values of these ratios at Earth 
for all values of $n$ from $0-\infty$, a range 
which covers all possible UHE neutrino flux sources. 
We show that the uncertainty on the 
predicted values of these flux ratios due to their 
dependence on the oscillation parameters threatens to wash out 
any sensitivity of the neutrino telescopes to $n$. 
We define a very simple $\chi^2$ function and show 
quantitatively the ranges of $n$ which could be ruled out 
by the data from UHE observations. 

We also point out, for the first time, a special mixing scenario 
for which the
observable ratio of muon neutrinos to the other neutrinos is
independent of the initial mixing scenario. This occurs when the
averaged survival probability for muon neutrinos takes its minimal
allowed value of $\frac 13$. 
It illustrates 
nicely that the ratio of muon neutrinos with the other flavors alone 
may not be a good discriminator of different sources.\\ 

Our paper is build up as follows: in Section \ref{sec:nutel} we
summarize the neutrino mixing phenomenology framework at neutrino
telescopes. Section \ref{sec:n} sees the discussion of several
neutrino sources whose common features lead to the parameterization 
in Eq.~(\ref{eq:main}). The peculiar mixing scenario with 
a constant muon neutrino flux with respect to the 
total flux is presented in Section \ref{sec:ext}. The 
experimental determination of $n$ is
investigated in Section \ref{sec:results}, before we 
sum up and conclude in Section \ref{sec:concl}.

\section{\label{sec:nutel}Neutrino Mixing and Neutrino Telescopes}
Astrophysical sources will generate fluxes of electron, 
muon and tau neutrinos, 
denoted by $\Phi_e^0$, $\Phi_\mu^0$ and $\Phi_\tau^0$, respectively. 
As a consequence 
of non-trivial neutrino mixing, it is not this initial flux 
composition which arrives at terrestrial detectors.  
In fact, what is measurable is given by\footnote{Everywhere in 
this paper we denote the fluxes at the source with a 
superscript 0 ($\Phi^0$), while 
the observed fluxes will be denoted without any superscript.} 
\be
\left( 
\ba
\Phi_e \\
\Phi_\mu \\
\Phi_\tau 
\ea
\right) 
= 
\left( 
\bad 
P_{ee} & P_{e \mu} & P_{e \tau} \\
P_{\mu e } & P_{\mu\mu} & P_{\mu \tau} \\
P_{\tau e} & P_{\tau \mu} & P_{\tau \tau} 
\ea
\right) 
\left( 
\ba
\Phi_e^0 \\
\Phi_\mu^0 \\
\Phi_\tau^0 
\ea
\right) ,
\ee
where the neutrino mixing probability is 
\be 
P_{\alpha \beta} = P_{\beta \alpha} = \sum\limits_i |U_{\alpha i}|^2\, 
|U_{\beta i}|^2~ 
\ee
and $U$ is the lepton mixing matrix.
The current best-fit values as well as the 
allowed $ 1\sigma$, $2\sigma$ and $3\sigma$ ranges of the 
oscillation parameters are \cite{limits} 
\be
\bad 
\sin^2 \theta_{12} & \sin^2 \theta_{13} 
& \sin^2 \theta_{23} \\ \hline
0.312 & 0.016 & 0.466 \\
0.294 \div 0.331 & 0.006 \div 0.026 & 0.408 \div 0.539 \\
0.278 \div 0.352 & 0.000 \div 0.036 & 0.366 \div 0.602 \\
0.263 \div 0.375 & 0.000 \div 0.046 & 0.331 \div 0.644 \\
\ea 
\ee
These mixing angles can be related to elements of the PMNS 
mixing matrix via 
\bea \label{eq:Upara}
U = 
\left( \bad 
c_{12} \, c_{13} & s_{12}   \, c_{13} & s_{13}  \, e^{-i \delta} \\[0.2cm] 
-s_{12}   \, c_{23} - c_{12}   \, s_{23}   \, s_{13}   \, e^{i \delta}  
& c_{12}   \,  c_{23} - s_{12}  \,   s_{23}  \,   s_{13}  \,  e^{i \delta}
& s_{23}   \,  c_{13}  \\[0.2cm] 
s_{12}  \,   s_{23} - c_{12}  \,   c_{23}  \,   s_{13}  \, e^{i \delta}& 
- c_{12}  \,   s_{23} - s_{12}  \,   c_{23}   \,  s_{13} \,    e^{i \delta}
& c_{23}   \,  c_{13}  \\ 
               \ea   \right)  , 
\eea
where $c_{ij} = \cos\theta_{ij}$, 
$s_{ij} = \sin\theta_{ij}$. 
The CP phase $\delta$ is unknown. Because the travelled distance of
the neutrinos is much larger than the oscillation length 
$4\pi \, E/\Delta m^2$, the mass-squared differences $\Delta m^2$
drop out of the mixing probabilities. Furthermore, 
solar neutrino mixing is neither maximal, nor 
zero or $\pi/2$, and due to these reasons no transition probability 
$P_{\alpha \beta}$ with $\alpha \neq \beta$ is zero 
and no survival probability $P_{\alpha \alpha}$ is one \cite{WR}. 
Consequently, high energy astrophysical neutrinos will always mix. 
To be precise, at $1\sigma$ and $3 \sigma$ the entries of the 
flavor conversion matrix $P_{\alpha \beta}$ are 
\be \label{eq:ranges}
P  = 
\left\{
\baz 
\left( 
\bad 
0.529 \div 0.578 & 0.178 \div 0.296 & 0.158 \div 0.275 \\
\cdot & 0.341 \div 0.443 & 0.354 \div 0.394 \\
\cdot & \cdot & 0.354 \div 0.469
\ea
\right) & (\mbox{at } 1\sigma)~,\\[0.3cm]
\left( 
\bad 
0.486 \div 0.612 & 0.127 \div 0.344 & 0.118 \div 0.335 \\
\cdot & 0.333 \div 0.508 & 0.304 \div 0.403 \\
\cdot & \cdot & 0.333 \div 0.525
\ea
\right) & (\mbox{at }3\sigma)~.
\ea
\right. 
\ee
The consequences and phenomenology of the special values 
$P_{\mu\mu} = 0.333$ and $P_{\tau\tau} = 0.333$ will be dealt 
with in Section \ref{sec:ext}. 

As is obvious from the ranges of the mixing parameters given above, 
a good zeroth order description of neutrino mixing is 
\be \label{eq:tbm}
U \simeq 
\left( 
\bad
\cos \theta_{12}  &  \sin \theta_{12}  & 0 \\[0.cm]
-\frac{\sin \theta_{12}}{\sqrt{2}} 
& \frac{\cos \theta_{12}}{\sqrt{2}}  & -\frac{1}{\sqrt{2}}\\[0.cm]
-\frac{\sin \theta_{12}}{\sqrt{2}} & 
\frac{\cos \theta_{12}}{\sqrt{2}}   & \frac{1}{\sqrt{2}}
\ea 
\right) .
\ee
Therefore, it proves in particular useful to expand 
in terms of 
\be 
|U_{e3}|\mbox{ and } 
\epsilon \equiv \frac{\pi}{4} - \theta_{23} \, .
\ee
Note that $\frac 12 - \sin^2 \theta_{23} = \epsilon 
+ {\cal O}(\epsilon^3)$. 
The result of the expansion for the flavor mixing matrix is 
\bea \label{eq:res1a}
P 
\simeq   
\left( 
\bad
 1 - 2 \, c_{12}^2 \, s_{12}^2 &  c_{12}^2 \, s_{12}^2  
& c_{12}^2 \, s_{12}^2    \\
 \cdot & \frac 12 \, (1 - c_{12}^2 \, s_{12}^2) &  
\frac 12 \, (1 - c_{12}^2 \, s_{12}^2)  \\
 \cdot & \cdot  & \frac 12 \, (1 - c_{12}^2 \, s_{12}^2)
\ea
\right) \\ 
+ \Delta 
\left( 
\bad
0 & 1 & -1 \\ 
\cdot & -1 & 0 \\ 
\cdot & \cdot & 1
\ea
\right) 
+ 
\frac 12 \, \overline{\Delta}^2
\left( 
\bad
0 & 0 & 0 \\ 
\cdot  & 1 & -1 \\ 
\cdot  & \cdot  & 1
\ea
\right) 
- \tilde{U}^2 
\left( 
\bad
2 & -1 & -1 \\ 
\cdot & \frac 12 & \frac 12 \\ 
\cdot & \cdot & \frac 12
\ea
\right) 
,
\ea
\ee
where the universal first \cite{xing,WR} and second \cite{PRW0,PRW1} 
order correction terms are 
\bea \D  \label{eq:Del}
\Delta = 
\frac 12 \, \sin^2 2 \theta_{12} \, \epsilon + 
\frac 14 \, \sin 4 \theta_{12} \, \cos \delta \, |U_{e3}| 
\simeq  \frac 19 
\left( 
\sqrt{2} \, \cos \delta \, |U_{e3}| + 4 \, \epsilon
\right)~,\\[0.2cm] \D 
\overline{\Delta}^2 =
3 \, \epsilon^2 + (\cos 2 \theta_{12} \, \epsilon - \sin 2 \theta_{12}
\, \cos \delta \, |U_{e3}| )^2 
\simeq 
3 \, \epsilon^2 + (2\sqrt{2} \, \cos \delta \, |U_{e3}| - \epsilon)^2 
 \, .
\eea
We have also given the expressions for the value 
$\sin^2 \theta_{12} = \frac 13$. 
Note that $\overline{\Delta}^2$ is positive semi-definite. The same
holds for the third expansion parameter  
\be \label{eq:tildeu}
\tilde{U}^2 = (1 - 2 \, c_{12}^2 \, s_{12}^2 ) \, |U_{e3}|^2 
\simeq \frac 59 \, |U_{e3}|^2  \,, 
\ee
which is usually only a sub-leading correction. 
The parameter $\overline{\Delta}^2$ can seldomly be larger than 
the first order term $\Delta$. To be quantitative, 
\be \label{eq:range2}
\bag   
\mbox{at } 1\sigma:& -0.046 \le \Delta \le 0.069 & \mbox{, } 
& \overline{\Delta}^2 \le 0.060  & \mbox{, } 
& 0.003 \le \tilde{U}^2 \le 0.015 ~,\\[0.2cm]
\mbox{at } 3\sigma:& -0.101 \le \Delta \le 0.112 & \mbox{, } & 
\overline{\Delta}^2 \le 0.162 & \mbox{, } 
& \tilde{U}^2 \le 0.028  ~.
\ea
\ee
The dependence of the expansion parameters 
on $\theta_{12}$ is weak. 
From the expressions for $\Delta$ and $\overline{\Delta}^2$ 
it is clear that their dependence on the atmospheric 
neutrino mixing angle $\theta_{23}$ is stronger than 
the one on $|U_{e3}| \, \cos \delta $ \cite{CNR}.

\section{\label{sec:n}Initial Flux Compositions and 
their Parametrization}

We now turn to the observable flavor ratios \cite{ratio1,ratio2}.  
The most frequently considered and experimentally most 
accessible is the ratio of flux of muon neutrinos to 
that of all other flavors: 
\be \label{eq:defT}
T = \frac{\Phi_\mu}{\Phi_e + \Phi_\mu + \Phi_\tau} 
= \frac{\Phi_\mu}{\Phi_{\rm tot}} \,.
\ee
Often one considers also the ratio 
$\Phi_\mu/(\Phi_e  + \Phi_\tau)$, which is simply 
$T/(1 - T)$. Whereas muon neutrinos with their characteristic 
tracks (continuous loss of energy via Cerenkov radiation) 
can be somewhat easily distinguished from electron and tau neutrinos, 
the distinction of the latter two is more difficult. 
If possible (e.g., via electromagnetic vs.~hadronic showers, 
or tau-induced lollipop/double bang events), one may consider 
their ratio as well: 
\be \label{eq:defR}
R = \frac{\Phi_e}{\Phi_\tau} \, .
\ee
We will use these ratios $T$ and $R$ in this paper. When neutrinos and
anti-neutrinos are not distinguished, then two independent ratios are
sufficient to fully determine the flavor content. For instance, if
both $T$ and $R$ are known, the ratio of muon neutrinos to tau
neutrinos $S$ is related to $T$ and $R$ via $T = S/(S + R + 1)$.

In what regards initial flux compositions, 
typically one considers pionic beam-dump-like sources, 
which have an initial flux composition of 
\be \label{eq:pi}
(\Phi_e^0 : \Phi_\mu^0 : \Phi_\tau^0) = (1 : 2 : 0) \, .
\ee
They arise because hadronic processes in a cosmic ray source produce 
pions. If the medium in such a source is opaque to muons,  
then one speaks of muon-damped cases \cite{010}, which have 
a composition of: 
\be \label{eq:md}
(\Phi_e^0 : \Phi_\mu^0 : \Phi_\tau^0) = (0 : 1 : 0) \, . 
\ee
Another possibility are so-called neutron beams, which result 
when photo-disintegration creates neutrons from 
nuclei \cite{100}, and therefore 
\be \label{eq:n}
(\Phi_e^0 : \Phi_\mu^0 : \Phi_\tau^0) = (1 : 0 : 0) \, .
\ee
Finally, at high energies semileptonic decays of charm 
quarks\footnote{For bottom quark decays 
the production rate is suppressed by one order of magnitude.} may
generate an initial flavor ratio of 
\cite{110}
\be \label{eq:charm}
(\Phi_e^0 : \Phi_\mu^0 : \Phi_\tau^0) = (1 : 1 : 0) \, .
\ee
We see that in all sources no tau neutrinos are generated initially.  
There is in fact always a small component (``prompt $\nu_\tau$'') 
from decays such as $D_s \ra \tau \, \nu_\tau$, however this 
contribution is at most at the permille level and can 
safely be neglected. 
We stress however that all of the above sources can be expected to 
be ``impure'', i.e., there will be small deviations from 
the idealized compositions given above \cite{lipari,5,PRW1}. 
As one example, Ref.~\cite{lipari} has argued that because of the 
wrong helicity polarization of the muons in pion decay the 
$\nu_\mu$ have actually a softer spectrum, and thus the effective
count of muon neutrinos is reduced. This depends on the injection
spectrum, which is described by the power law $E^{-\alpha}$. For
the canonical value $\alpha = 2$ it was found that effectively 
$(\Phi_e^0 : \Phi_\mu^0 : \Phi_\tau^0) = (1 : 1.86 : 0)$. It was 
shown \cite{PRW1} that inclusion of leptonic and semi-leptonic 
kaon decays, as well as of heavy flavor decays does not change this
ratio appreciably. 

One may also consider GZK-neutrinos \cite{gzk}, whose flavor content
changes with energy. Below about 100 PeV ($10^8$ GeV) one has 
$(1 : 0 : 0)$, whereas for higher energies $(1 : 2 : 0)$.\\

Given all the above discussion, we are lead to parameterize the 
initial flux composition of high energy neutrinos simply as 
\be  \label{eq:1n1}
(\Phi_e^0 : \Phi_\mu^0 : \Phi_\tau^0) = (1 : n : 0) \, .
\ee
The experimental determination of the initial flux composition 
is therefore simply an extraction of the parameter $n$ from the 
observed flux composition on Earth\footnote{Ref.~\cite{zz} has 
proposed 
$(\Phi_e^0 : \Phi_\mu^0 : \Phi_\tau^0) = (\sin^2 \xi \, \cos^2 \zeta 
: \cos^2 \xi \, \cos^2 \zeta :  \sin^2 \zeta)$.}. 
From Eq.~(\ref{eq:1n1}) one can obtain (pure) pion sources 
when $n = 2$, 
neutron beams for $n = 0$, charm sources for $n = 1$ 
and muon-damped sources when one takes the limit 
$n \rightarrow \infty$.

Let us attempt to express the flux ratios $T$ and $R$ 
in a source-independent way. We take advantage of the 
parametrization in Eq.~(\ref{eq:1n1}) and use the expansion of the 
probabilities $P_{\alpha\beta}$ in Eq.~(\ref{eq:res1a}). 
First we consider the ratio of muon to
all neutrinos and find 
\bea \label{eq:Tgen}\D 
T = \frac{\Phi_\mu}{\Phi_{\rm tot}} = 
\frac{P_{e \mu} + n \, P_{\mu \mu}}{1 + n} \\ \D 
\simeq \frac{1}{1 + n} 
\left[
\frac n2 + \left(1 - \frac n2\right) \, c_{12}^2 \, s_{12}^2 + 
\Delta \, \left(1 - n\right) + \frac n2 \, \overline{\Delta}^2 + 
\tilde{U}^2 \, \left(1 - \frac n2\right)
\right]\,.
\eea
\begin{table}[th]
\begin{center}
\begin{tabular}{c|c|c|c||c|c}
& $n$ & \multicolumn{2}{c||}{$T$} 
& \multicolumn{2}{c}{$R$} \\ \hline 
& & $1\sigma$ & $3\sigma$ & $1\sigma$ & $3\sigma$ \\ \hline \hline
pion & 2 & $0.324 \div 0.355$ & $0.323 \div 0.387$ 
& $0.893 \div 1.247$ & $0.820 \div 1.450$  \\ \hline
charm & 1 & $0.303 \div 0.323$ & $0.298 \div 0.348$ 
& $1.121 \div 1.581$ & $0.998 \div 1.853$  \\ \hline
neutron & 0 & $0.178 \div 0.296$ & $0.127 \div 0.345$ 
& $1.922 \div 3.618$ & $1.449 \div 5.090$  \\ \hline
muon-damped & $\infty$ & $0.341 \div 0.443$ & $0.333 \div 0.508$ 
& $0.462 \div 0.814$ & $0.344 \div 1.071$  \\ \hline
\end{tabular}
\caption{\label{tab:ranges}Ranges of the flux ratios 
$T = \Phi_{\mu}/\Phi_{\rm tot}$ and $R = \Phi_{e}/\Phi_{\tau}$ 
for an initial flux composition of 
$(\Phi_e^0 : \Phi_\mu^0 : \Phi_\tau^0) = (1 : n : 0)$. 
The case $n = 2$ correspond to pion sources, $n = 1$ to charm 
sources, $n = 0$ to neutron beams and $n \rightarrow \infty$ to 
muon-damped sources. We have inserted the $1\sigma$ and 
$3\sigma$ ranges of the oscillation parameters. }
\end{center}
\end{table}
We note the following points from this expression: 
\begin{itemize}
\item for $n=2$, which corresponds 
to pionic sources, there is no explicit dependence on 
$\theta_{12}$ and also corrections due to $\tilde{U}^2$ vanish. 
The uncertainty on $T$ is hence expected to 
be amongst the lowest for this case; 
\item for $n=1$, which corresponds 
to charm sources, there is no first order correction $\Delta$, and
thus corrections to the zeroth order expression are 
only of quadratic order in the small expansion parameters 
$\theta_{23} - \pi/4$ and $|U_{e3}|$; 
\end{itemize}
In the zeroth order approximation, which corresponds to 
$\mu$--$\tau$ symmetry yielding $\theta_{13}=0$ 
and $\theta_{23}=0$ such that 
the correction terms $\Delta$, $\overline{\Delta}$ and $\tilde{U}$ 
are all vanishing, Eq.~(\ref{eq:Tgen}) reduces to 
\[
T = \frac{1}{1 + n} 
\left[\frac n2 + \left(1 - \frac n2\right) 
\, c_{12}^2 \, s_{12}^2\right] 
\rightarrow \left\{ 
\bad 
\frac 13 & \mbox{for } n = 2 & \mbox{(pion source)}\\
\frac 14 \, (1 + c_{12}^2 \, s_{12}^2 ) \simeq \frac{11}{36} 
& \mbox{for } n = 1 & \mbox{(charm source)}\\
c_{12}^2\, s_{12}^2 \simeq \frac 29 & 
\mbox{for } n = 0 & \mbox{(neutron beam)}\\
\frac 12 \, (1 - c_{12}^2
\, s_{12}^2 ) \simeq \frac{7}{18} 
& \mbox{for } n = \infty & \mbox{(muon-damped)}\\
\ea
\right. \, ,
\]
where we have also given the simple expressions for 
the special values of $n$ corresponding to the usual sources one
considers. 
Hence, we note that all sources have similar predicted $T$ 
around 0.3. Indeed, in Section \ref{sec:ext} we
will present an extreme scenario in which $T = \frac 13$ independent
of $n$. We further note that the predicted $T$ is highest for 
$n=\infty$, lowest for $n=0$ and 
intermediate for $n=1$ and 2. 
In general, $T$ slightly increases with $n$.

Similarly, the ratio $R$ of electron and tau neutrinos is 
\bea \label{eq:Rgen}\D 
R = \frac{\Phi_e}{\Phi_{\tau}} = 
\frac{P_{e e} + n \, P_{e \mu}}{P_{e \tau} + n \, P_{\mu \tau}} 
\\ \D
= \frac{1 - 2 \, c_{12}^2 \, s_{12}^2 \left(1 - \frac n2 \right) 
+ n \, \Delta - 2 \, \tilde{U}^2 \, \left(1 - \frac n2\right)}
{\frac n2 + c_{12}^2 \, s_{12}^2 \left(1 - \frac n2 \right) - \Delta 
- \frac n2 \, \overline{\Delta}^2 + 
\tilde{U}^2 \, \left(1 - \frac n2\right)} \, .
\eea
Here the value $n = 2$ removes again several correction terms, in
particular the terms with explicit dependence on $\theta_{12}$, 
which leads to small uncertainty.  
For the case of $\mu$--$\tau$ symmetry one finds
\[
R = \frac{1 - 2 \, c_{12}^2 \, s_{12}^2}
{\frac n2 + c_{12}^2 \, s_{12}^2 \, (1 - \frac n2)} 
\rightarrow \left\{ 
\bad 
1 & \mbox{for } n = 2 & \mbox{(pion source)}\\
2\,\frac{1 - c_{12}^2 \, s_{12}^2}
{1 - c_{12}^2 \, s_{12}^2} \simeq \frac{14}{11} 
& \mbox{for } n = 1 & \mbox{(charm source)}\\
\frac{1 - 2 \, c_{12}^2 \, s_{12}^2}
{c_{12}^2 \, s_{12}^2} \simeq \frac 52 & 
\mbox{for } n = 0 & \mbox{(neutron beam)}\\
2 \,\frac{c_{12}^2 \, s_{12}^2}{1 - c_{12}^2 \, s_{12}^2} 
\simeq \frac{4}{7} 
& \mbox{for } n = \infty & \mbox{(muon-damped)}\\
\ea
\right. \, .
\]
The spread for the different values of $n$ is larger than for 
$T$. For instance, if $n = 2$ then $R = 1$ at leading order and 
between $0.820 \div 1.450$ for the $3\sigma$ range, while 
$n = 1$ leads to $R = \frac{14}{11}$ at leading order and 
between $0.998 \div 1.853$ for the $3\sigma$ range. 
In general, $R$ slightly decreases with $n$.

\begin{figure}
\begin{center}
\includegraphics[width=0.58\textwidth,angle=270]{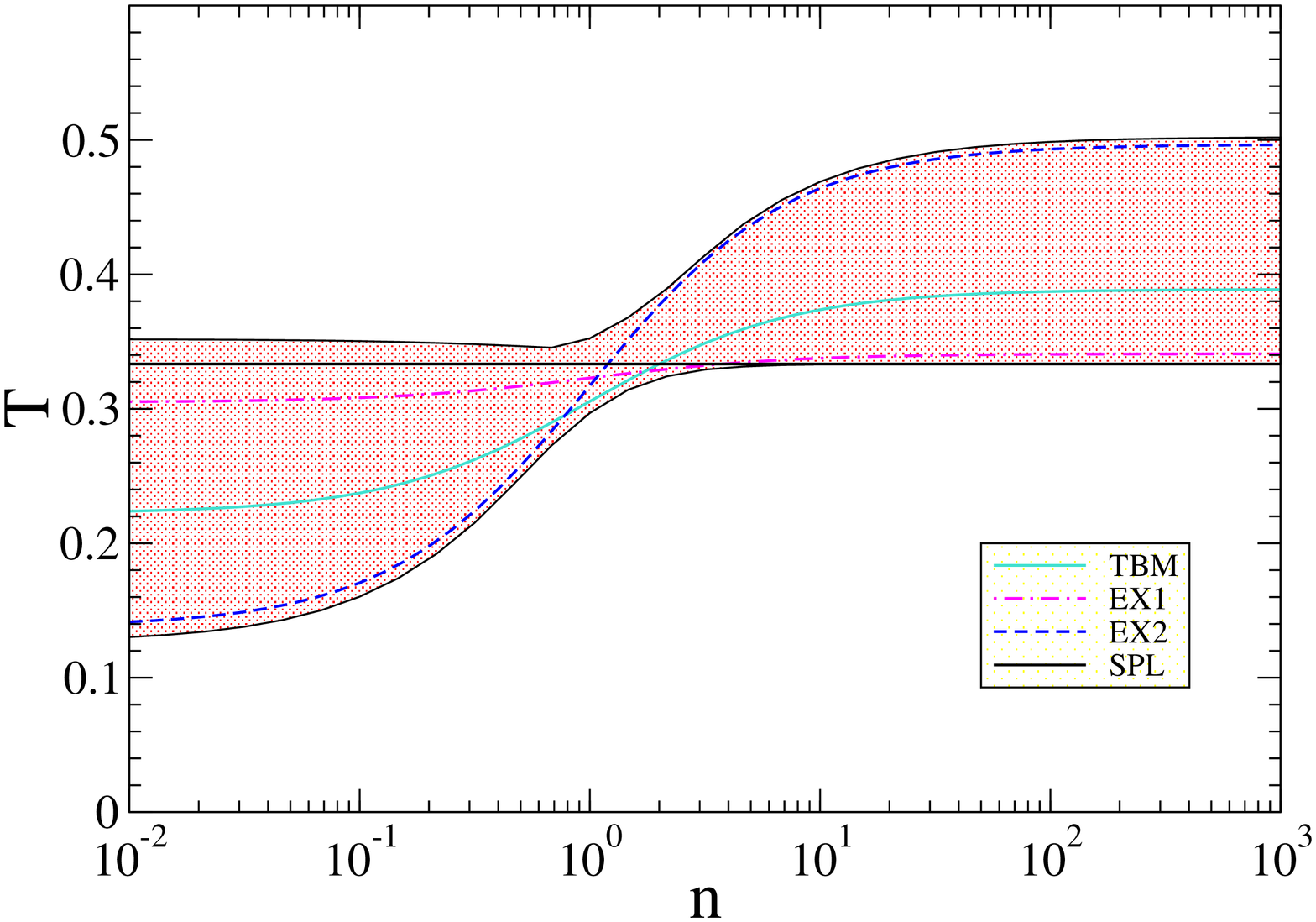}
\includegraphics[width=0.58\textwidth,angle=270]{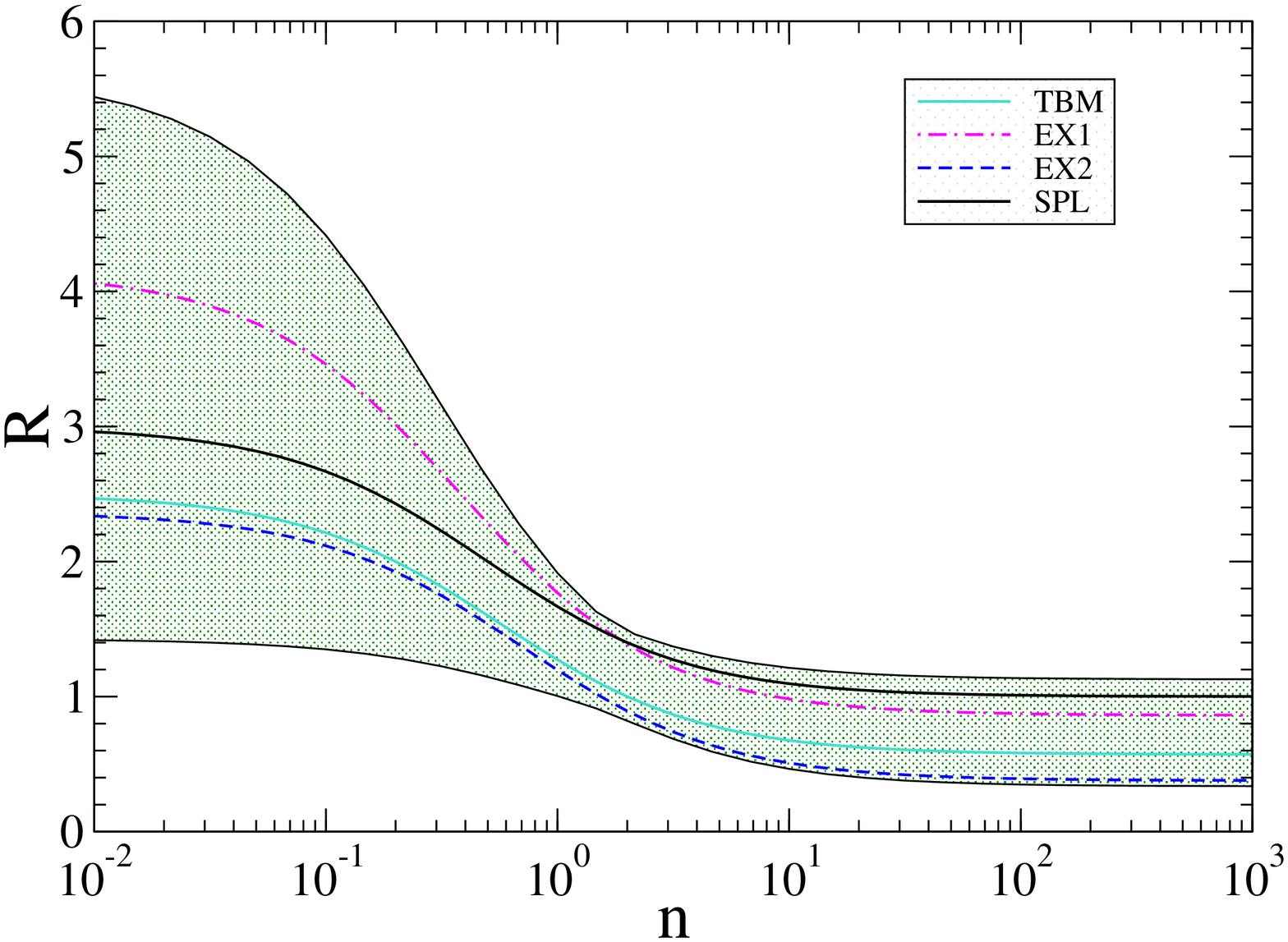}
\caption{\label{fig:TR}
Range of allowed values of 
$T = \Phi_\mu/\Phi_{\rm tot}$ (upper plot) and 
$R =  \Phi_e/\Phi_{\tau}$ (lower plot) 
as a function of the flux composition 
parameter $n$, when the oscillation parameters are varied
within their $3\sigma$ ranges. }
\end{center}
\end{figure}

In Fig.~\ref{fig:TR} we show 
$T$ and $R$ as a function of $n$. 
In Table \ref{tab:ranges} we present the 1$\sigma$ and 
3$\sigma$ ranges of predicted $T$ and $R$ for the 
4 benchmark cases of $n$ 
corresponding to pion, charm, neutron and muon-damped sources. 
The 1$\sigma$ and 3$\sigma$ ranges of $T$ and $R$ correspond to 
current 1$\sigma$ and $3\sigma$ ranges of 
the oscillation parameters. 
In general one notes again that $T$ is not an ideal discriminator 
in order to distinguish the neutrino source, because it is rather
close to $\frac 13$ for all $n$. Therefore, at the present 
stage we can already suspect that 
the task of identifying $n$ will be easier when $R$ is added to the 
statistics. A more detailed analysis will be presented in Section 5.

\section{\label{sec:ext}An Extreme Case: Minimal Survival Probabilities}

One observes from the allowed ranges of the 
oscillation probabilities in Eq.~(\ref{eq:ranges}) 
that $P_{\mu\mu}$ and $P_{\tau\tau}$ can take the value $0.333$ 
when the $3\sigma$ ranges of the oscillation parameters 
are inserted. It is worth noting that $\frac 13$ is the minimal value
that an averaged survival probability $P_{\alpha\alpha}$ can take if
there are 3 fermion families. In general, in the presence of 
$k$ flavors we have 
\be
P_{\alpha\alpha}^{\rm min} = \frac 1k \, .
\ee
The minimum is obtained if and only if  
$|U_{\alpha i}|^2 = 1/k$ for all $i$, i.e., when all mixing 
matrix elements of a row take on the same value\footnote{ 
Note that this concerns the elements of
a {\it row}. Commonly it is assumed the elements of a column are
identical (as for tri-bimaximal, or more general 
tri-maximal mixing \cite{tm}).}.  

\begin{figure}
\begin{center}
\epsfig{file=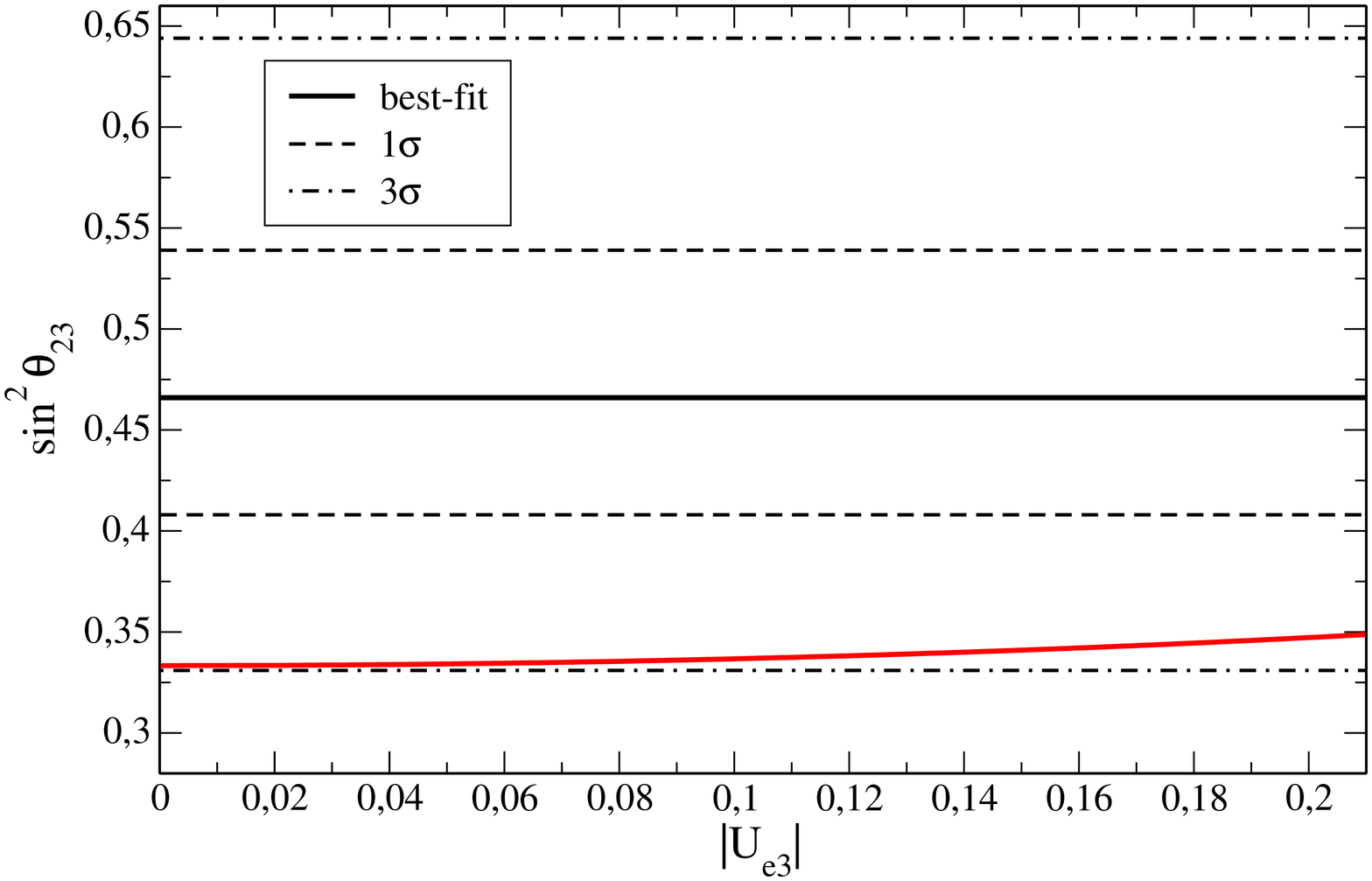,width=10cm,height=7cm}
\epsfig{file=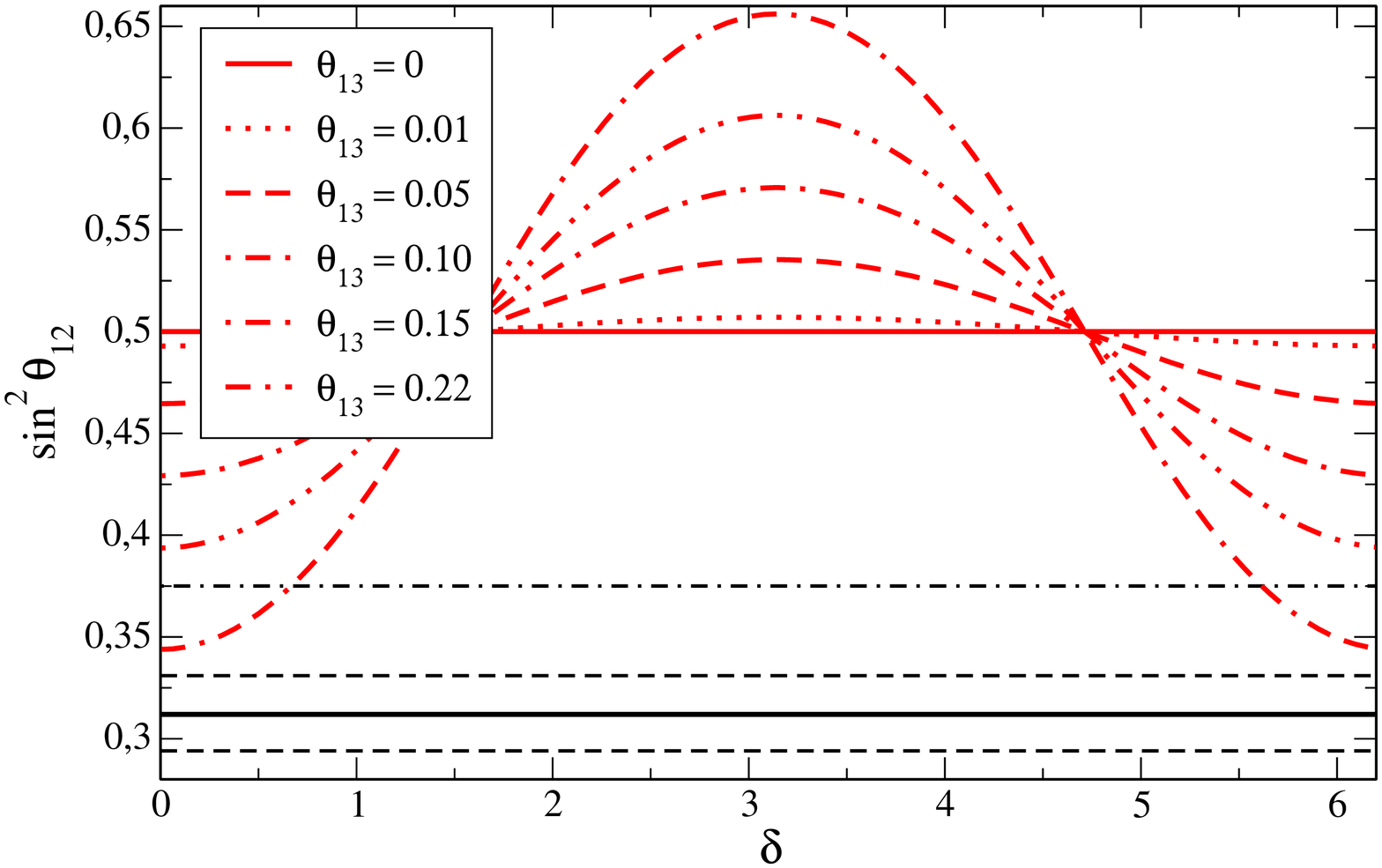,width=10cm,height=7cm}
\epsfig{file=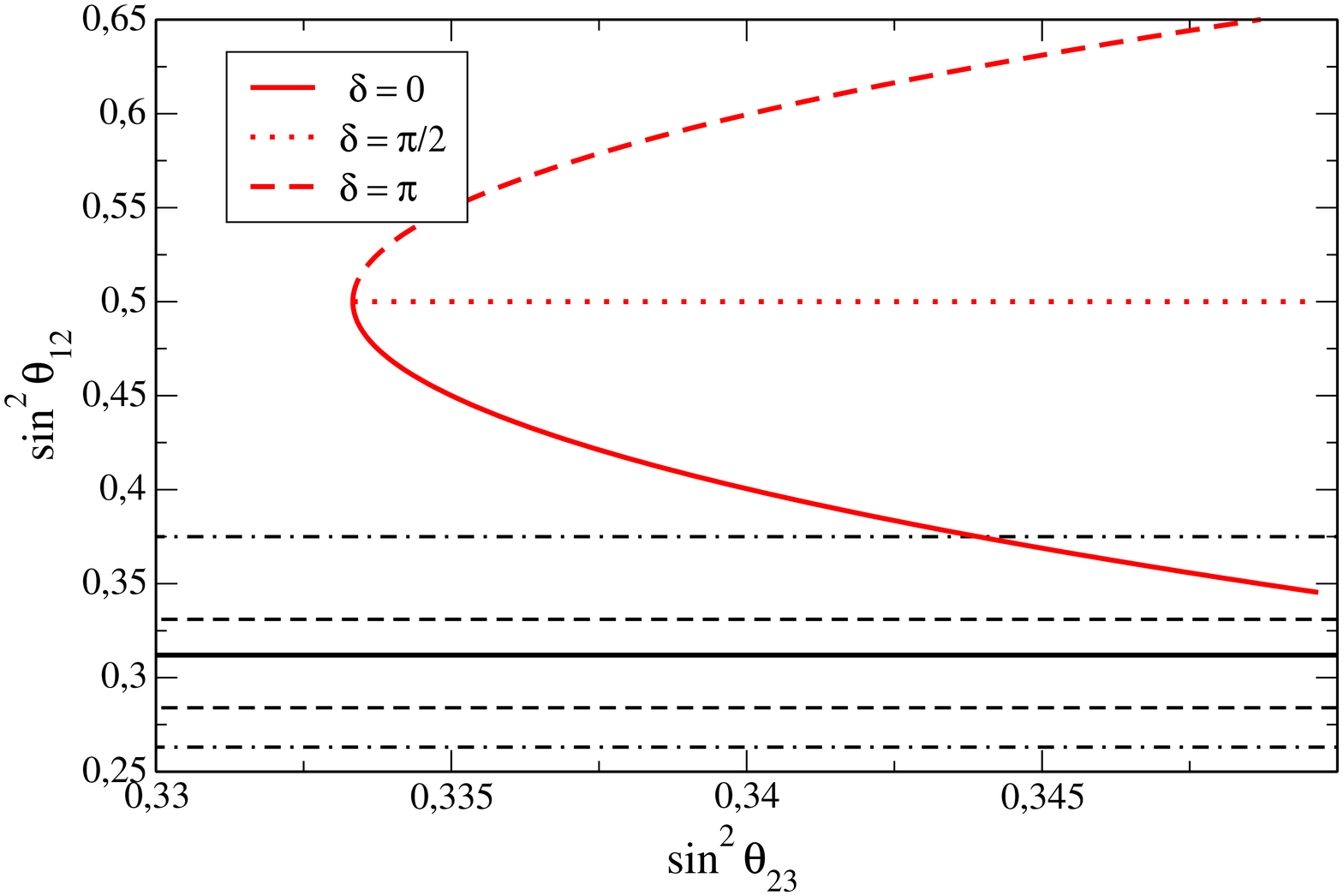,width=10cm,height=7cm}
\caption{\label{fig:m}
Phenomenology of a PMNS matrix with all elements of the second row 
equal to each other. 
Shown are the atmospheric neutrino parameter $\sin^2
\theta_{23}$ against $|U_{e3}|$, the solar neutrino 
parameter $\sin^2 \theta_{12}$ against $\delta$ for 
different values of $|U_{e3}|$ and $\sin^2
\theta_{12}$ against $\sin^2 \theta_{23}$. 
Also given are the current best-fit value and the 
$1\sigma$ as well as $3\sigma$ ranges of the oscillation 
parameters. } 
\end{center}
\end{figure}

Considering the case of $k = 3$, let us 
discuss first the phenomenological consequences of 
$|U_{\mu i}| = \sqrt{1/3}$. The case of $|U_{\tau i}| = \sqrt{1/3}$ will
be discussed at the end of this Section. 
First we note that 
unitarity  of the PMNS matrix links the three entries of a row, 
which implies that only two independent constraints can result. 
The first one stems from $|U_{\mu 3}| = \sqrt{1/3}$ and is 
\be \label{eq:cond1}
\sin^2 \theta_{23} = \frac 13 \frac{1}{1 - |U_{e3}|^2} \simeq 
\frac 13 \left(1 +|U_{e3}|^2 \right).
\ee
Inserting this in $|U_{\mu 1}| = \sqrt{1/3}$ gives the second
independent constraint 
\be \label{eq:cond2}
\cos \delta \, 
\tan 2 \theta_{12} = 
\frac{1 - 2 \, |U_{e3}|^2}{|U_{e3}| \, \sqrt{2 - 3 \, |U_{e3}|^2}}
\simeq \frac{1}{\sqrt{2}} \left( \frac{1}{|U_{e3}|} + 
\frac 54 \, |U_{e3}| \right)
 \,.
\ee
We immediately see that $|U_{e3}|$ should be rather large,  
$\sin^2 \theta_{23}$ significantly less than $\frac 12$ and 
$\cos \delta$ should be different from zero, or $\sin \delta$
different from 1. We display the phenomenology resulting from 
Eqs.~(\ref{eq:cond1}, \ref{eq:cond2}) in Fig.~\ref{fig:m}. 
While $\sin^2 \theta_{23}$ lies close to its lower $3\sigma$ bound, 
$\sin^2 \theta_{12}$ can approach its upper $1\sigma$ bound from
above, in which case $|U_{e3}|$ is close to its current upper bound
and $\delta$ less than one (or larger than 5). Vanishing 
$\theta_{13}$ leads to maximal $\theta_{12}$ and is disallowed.

The condition $|U_{\mu i}|^2 = \frac 13$ 
together with unitarity of the PMNS matrix fix the form 
of the matrix 
$|U_{\alpha i}|^2$, namely 
\be \label{eq:U2ex}
|U_{\alpha i}|^2 = 
\left(
\bad
|U_{e1}|^2 & 1 - |U_{e1}|^2 - |U_{e3}|^2 & |U_{e3}|^2 \\
\frac 13 & \frac 13 & \frac 13 \\
\frac 23 - |U_{e1}|^2 & |U_{e1}|^2 + |U_{e3}|^2 - \frac 13 & 
\frac 23 - |U_{e3}|^2
\ea
\right) .
\ee
Recall that the $|U_{\alpha i}|^2$, and therefore the averaged 
probabilities $P_{\alpha\beta}$, depend on four independent
parameters. The condition $|U_{\mu i}| = \sqrt{1/3}$ fixes two of the four
parameters and we have chosen 
$|U_{e1}|^2$ and $|U_{e3}|^2 $ as the remaining two free 
and independent parameters. It is easy to see that from 
Eq.~(\ref{eq:U2ex}) the relation 
\be
P_{\mu \alpha} = P_{\alpha \mu} 
= \frac 13 ~~\mbox{for }\alpha = e, \mu, \tau 
\ee
follows. With this relation one obtains a remarkably simple 
result for the ratio $T$ as a function of $n$, namely   
\be
T = \frac{\Phi_\mu}{\Phi_{\rm tot}} 
= \frac{P_{e\mu} + n \, P_{\mu\mu}}{1 + n} = \frac{1}{3} \, ,
\ee
regardless of $n$! 
The origin of this simple result is of course that all 
neutrinos $\nu_i$ have an equal share in $\nu_\mu$, i.e., their 
muon content is equally distributed\footnote{Similarly, the ratio of 
muon neutrinos to the sum of electron and tau neutrinos is 
$\Phi_\mu/(\Phi_e + \Phi_\tau) = \frac 12$.}. 
Concerning $R$, the general result as a function of $n$ is 
rather lengthy. The allowed ranges of $R$ are however 
significantly less than for the unconstrained case. 
For instance, $R$ for pion sources lies 
between 1.23 and 1.40, to be compared with the general 
allowed range between 0.82 and 1.45. An interesting case occurs for 
muon-damped sources, for which we find that $R = 1$.\\

If the third row of $U$ has identical entries, and therefore 
$P_{\tau\tau} = \frac 13$, then 
\be
\sin^2 \theta_{23} = \frac 13 \frac{1}{1 - |U_{e3}|^2} \, 
\left(2 - 3 \,|U_{e3}|^2  \right) 
\simeq \frac 23 \left(1 - \frac 12 \, |U_{e3}|^2 \right)
\ee
and 
\be
\cos \delta \, 
\tan 2 \theta_{12} = -
\frac{1 - 2 \, |U_{e3}|^2}{|U_{e3}| \, \sqrt{2 - 3 \, |U_{e3}|^2}}
\simeq \frac{-1}{\sqrt{2}} \left( \frac{1}{|U_{e3}|} -
\frac 54 \, |U_{e3}| \right)
 \,.
\ee
While the second constraint is very similar to the one for 
$|U_{\mu i}|^2 = \frac 13$, the first condition implies here 
that $\sin^2 \theta_{23}$ lies slightly outside its 
allowed $3\sigma$ range. The value $P_{\tau\tau}^{\rm min} = 0.333$ 
found in Eq.~(\ref{eq:ranges}) is therefore only numerically close to 
$\frac 13$. In any case, since the identification of muon neutrinos is
much simpler than of tau neutrinos, it makes little sense to discuss
the scenario with a saturated $P_{\tau\tau}^{\rm min}$.

\section{\label{sec:results}Experimental Distinction of Sources}

\begin{figure}
\begin{center}
\includegraphics[width=9.0 cm,height=16.0cm,angle=270]{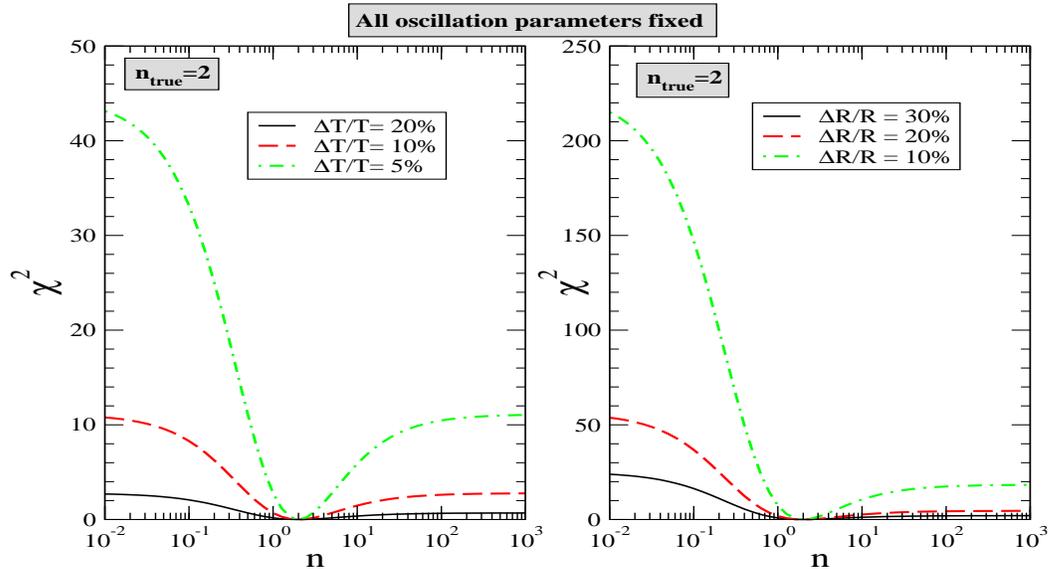}
\caption{\label{fig:ntruefixed}
$\chi^2$ as a function of $n_{fit}$ for $n_{true}=2$, for 
different values of errors on the flux ratios $T$ 
(left panel) and $R$ (right panel). All oscillation parameters 
are fixed in the fit. 
}
\end{center}
\end{figure}

In this Section we  
explore the potential of the neutrino telescopes to 
ascertain the initial flux composition at the source. 
We stress that all results to be shown have been obtained 
by an exact calculation of the oscillation probabilities. 
We begin by returning to Fig.~\ref{fig:TR}, in which we display 
the current $3\sigma$ predicted ranges of $T$ and $R$ 
respectively, as a function of the flux composition 
parameter $n$. For each value of $n$ we calculated the 
full range of predicted values of $T$ and $R$ by allowing 
the oscillation parameters to vary in their current 
$3\sigma$ allowed range \cite{limits}. The range of 
values of $T$ and $R$ for the different limiting cases 
are given in Table \ref{tab:ranges}. As discussed before, 
we note from the figure that the $T$ ranges for all $n$ 
are overlapping. We also note that 
the spread in $T$ is minimized for values of $n$
close to 2, as for this case the explicit dependence on 
$\theta_{12}$ vanishes (cf.~Eq.~(\ref{eq:Tgen})).  
For the same reason, the spread in $R$ is 
also minimized for values of $n$
close to 2 (cf.~Eq.~(\ref{eq:Rgen})). 
The spread in $T$ for small $n$ is somewhat larger
than for large $n$. 
The spread in $R$ for small $n$ is seen to be significantly 
larger than for large $n$. The large spread for small $n$ of both $T$
and $R$ implies that neutron sources have the largest dependence on
the neutrino mixing parameters. Note that while the 
possible 
spread induced by  neutrino mixing is smallest for $n$ between 
1 and 2, it does not necessarily imply that the sensitivity to 
initial flux composition will be best for these cases. 
We also note that $R$ appears to 
be a much better discriminator for $n$ compared to $T$ and hence  
inclusion of $R$ in addition to 
$T$ will make the identification of the source easier. The 
individual lines embedded in the band in Fig.~\ref{fig:TR} 
show the predicted $T$ and $R$ for 
several benchmark mixing scenarios, in which the
values of the oscillation parameters are 
\begin{itemize}
\item[``TBM''] the tri-bimaximal scenario: 
the oscillation parameters are 
$\sin^2 \theta_{12} = \frac 13$, $\sin^2 \theta_{23} = \frac 12$, 
$|U_{e3}|^2 = 0$ and $\delta=0$; 
\item[``EX1''] extreme scenario 1: the oscillation parameters are 
$\sin^2 \theta_{12} = 0.27$, $\sin^2 \theta_{23} = 0.35$, 
$|U_{e3}|^2 = 0.04$ and $\delta=0$; 
\item[``EX2''] extreme scenario 2: the oscillation parameters are 
$\sin^2 \theta_{12} = 0.27$, $\sin^2 \theta_{23} = 0.65$, 
$|U_{e3}|^2 = 0.0$ and $\delta=0$;
\item[``SPL''] special scenario: for the special scenario discussed  
in the previous Section, we take $\sin^2 \theta_{12} = 0.352$, 
$\sin^2 \theta_{23} = 0.348$, 
$|U_{e3}|^2 = 0.043$ and $\delta=0$.
\end{itemize}
One can immediately see from Fig.~\ref{fig:TR} 
that as predicted in Section \ref{sec:ext}, the 
special case yields $T=1/3$ for all values of $n$. 
Therefore, if this case was the true mixing scenario chosen 
by Nature, then it would be impossible to conclude anything 
about $n$ from $T$ measurement at neutrino telescopes. 
One can also 
note that $T$ for the EX1 mixing case is very close to 
the SPL case and hence it would also be difficult to say 
anything about the source, if this set of mixing parameters 
turn out to be the true 
mixing angles. 

\begin{figure}[t]
\begin{center}
\includegraphics[width=9.0 cm,height=7.0cm,angle=270]{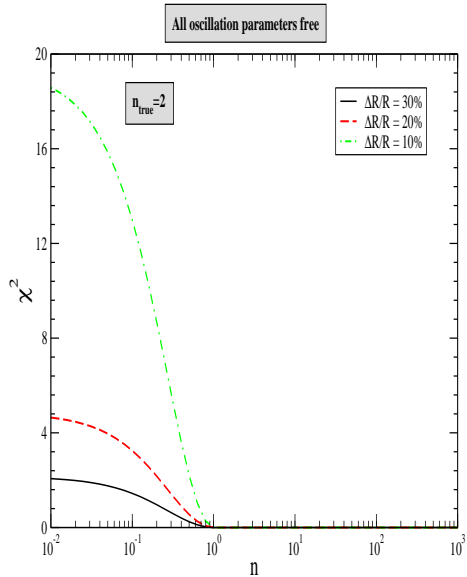}
\caption{\label{fig:ntruefree}
$\chi^2$ as a function of $n_{fit}$ for $n_{true}=2$, for 
different values of errors on the flux ratios 
$R$. All oscillation parameters 
allowed to vary freely within their current $3\sigma$ limits 
in the fit. Corresponding $\chi^2$ for $T$ is found to be zero 
for all values of $n_{fit}$ and all errors $\Delta T/T$.}
\end{center}
\end{figure}

We next define a very simple $\chi^2$ function 
as\footnote{We are aware 
that this definition of $\chi^2$ is not strictly valid for 
a ratio as its error will not be Gaussian. Nonetheless we use 
it here for the sake of simplicity as the purpose of this paper is 
to roughly illustrate the potential of the neutrino telescope 
rather than to show exact numerical results for which one should 
work with the actual number of events calculated using  a detailed 
code for the detector including threshold, efficiencies and so on.}
\be
\chi^2 = \left(\frac{F_{Data} - F_{ Theory}}{\sigma_F}\right)^2,
\label{eq:chisq}
\ee
where $F$ can be either $T$ or $R$ and ``$Data$'' 
and ``$Theory$'' refer to 
observed and predicted flux ratios respectively, and 
$\sigma_F$ is the 1$\sigma$ experimental error on the 
relevant flux ratio. We show results for fixed experimental errors 
of 20\%, 10\% and 5\% for $T$ and 30\%, 20\% and 10\% for $R$. 
We generate the ``$F_{Data}$'' at certain ``true'' values $n$, denoted as 
$n_{true}$, and at one of the 
benchmark mixing scenarios given above. This 
$F_{Data}$ is then fitted with $F_{Theory}$ corresponding to $n_{fit}$. 
The oscillation parameters in the fit are generally allowed to 
vary freely in the fit, apart from Fig. \ref{fig:ntruefixed}. 
\\

In Fig.~\ref{fig:ntruefixed} we show the $\chi^2$ obtained 
as a function of $n_{fit}$ for $n_{true}=2$ and TBM mixing. 
All oscillation parameters are kept fixed at their TBM values 
in the fit for this figure. 
We reiterate that  
$n_{true}$ is defined 
as the value of $n$ for which the $T_{Data}$ 
($R_{Data}$) are generated 
and $n_{fit}$ is the value of $n$ in $T_{Theory}$ ($R_{Theory}$). 
The best-fit comes at $n_{fit}=2$, as expected. We can see that 
extremely good sensitivity to $n$ comes for $\Delta T/T=5\%$, while 
reasonable sensitivity to $n$ is expected if $\Delta T/T=10\%$. 
For $\Delta T/T=20\%$ one finds hardly any sensitivity at all, 
even in this ideal case, where the effect of oscillation 
parameter uncertainties have been neglected. If $R$ could be 
measured at the neutrino telescope, then we expect  
good sensitivity even if $\Delta R/R=30\%$. 
\\

\begin{figure}[ht]
\begin{center}
\includegraphics[width=6cm,height=7cm,angle=270]{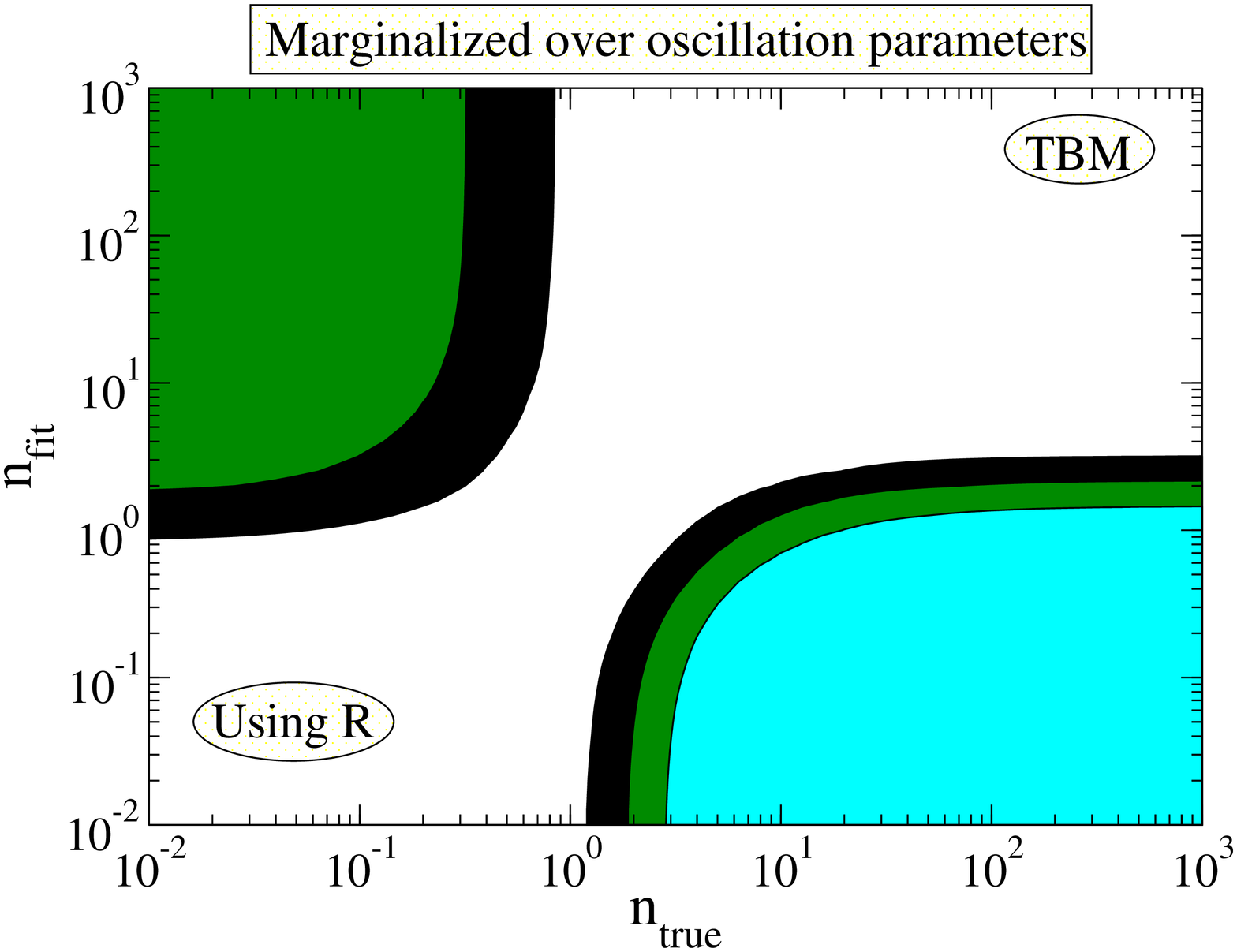}
\includegraphics[width=6cm,height=7cm,angle=270]{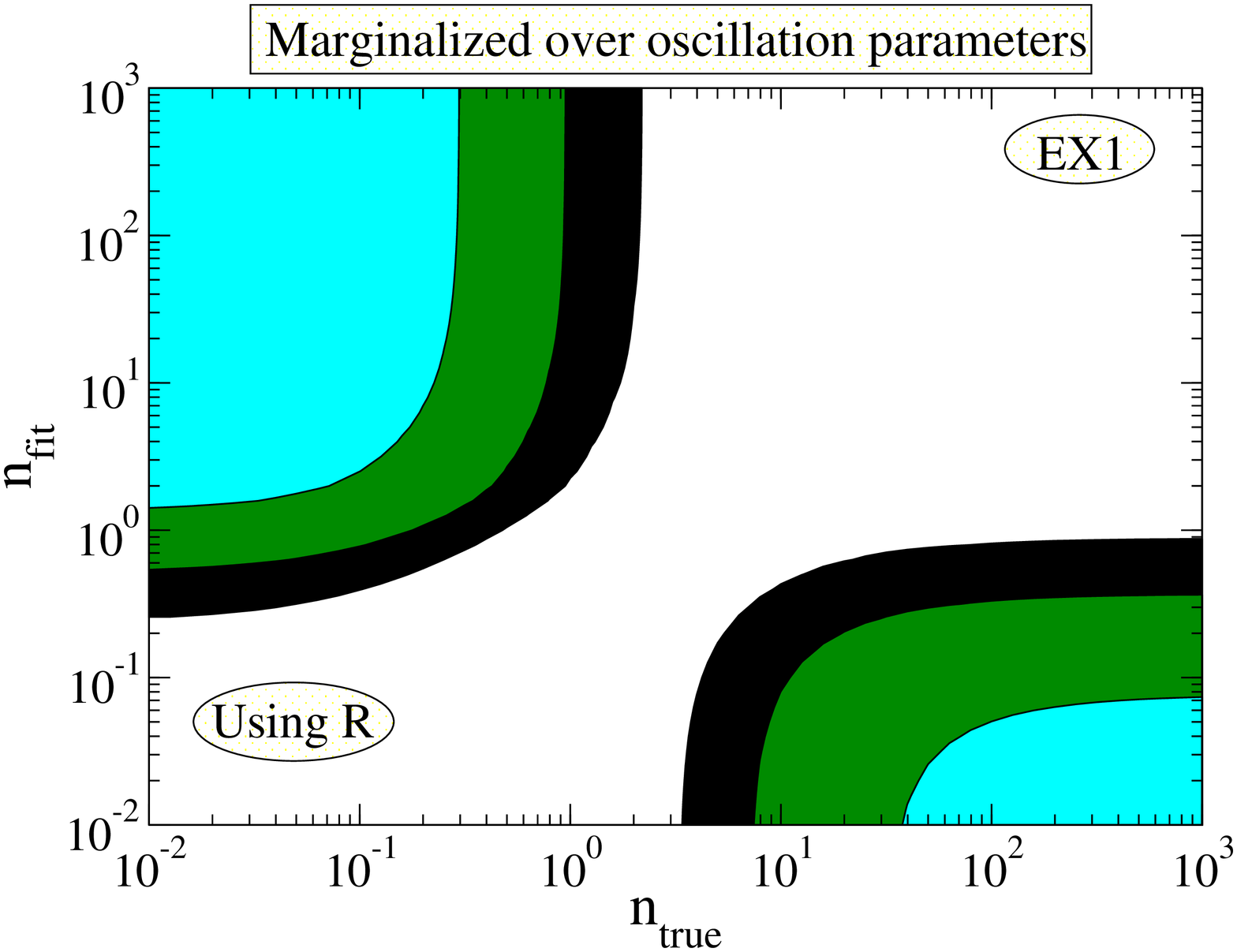}

\includegraphics[width=6cm,height=7cm,angle=270]{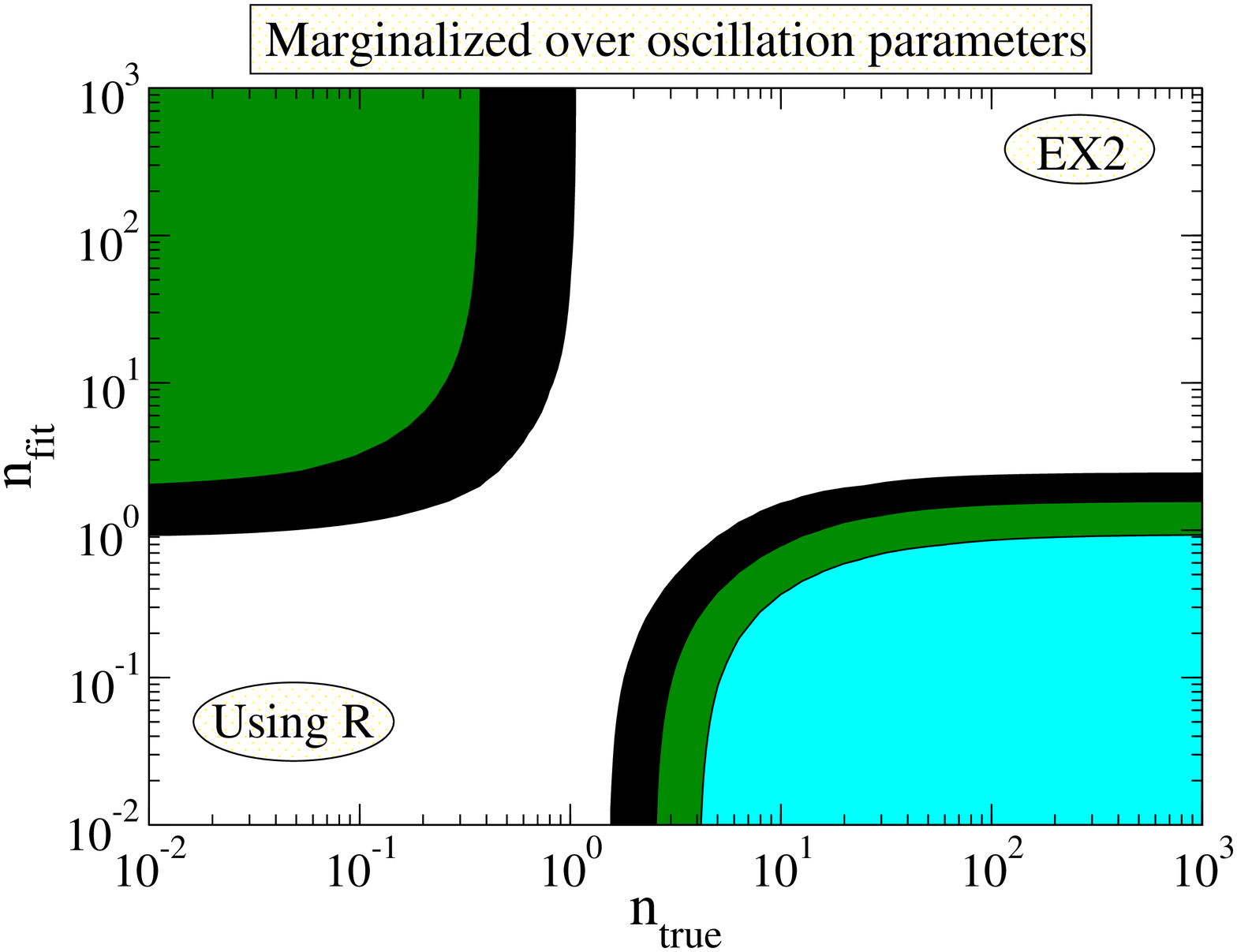}
\includegraphics[width=6cm,height=7cm,angle=270]{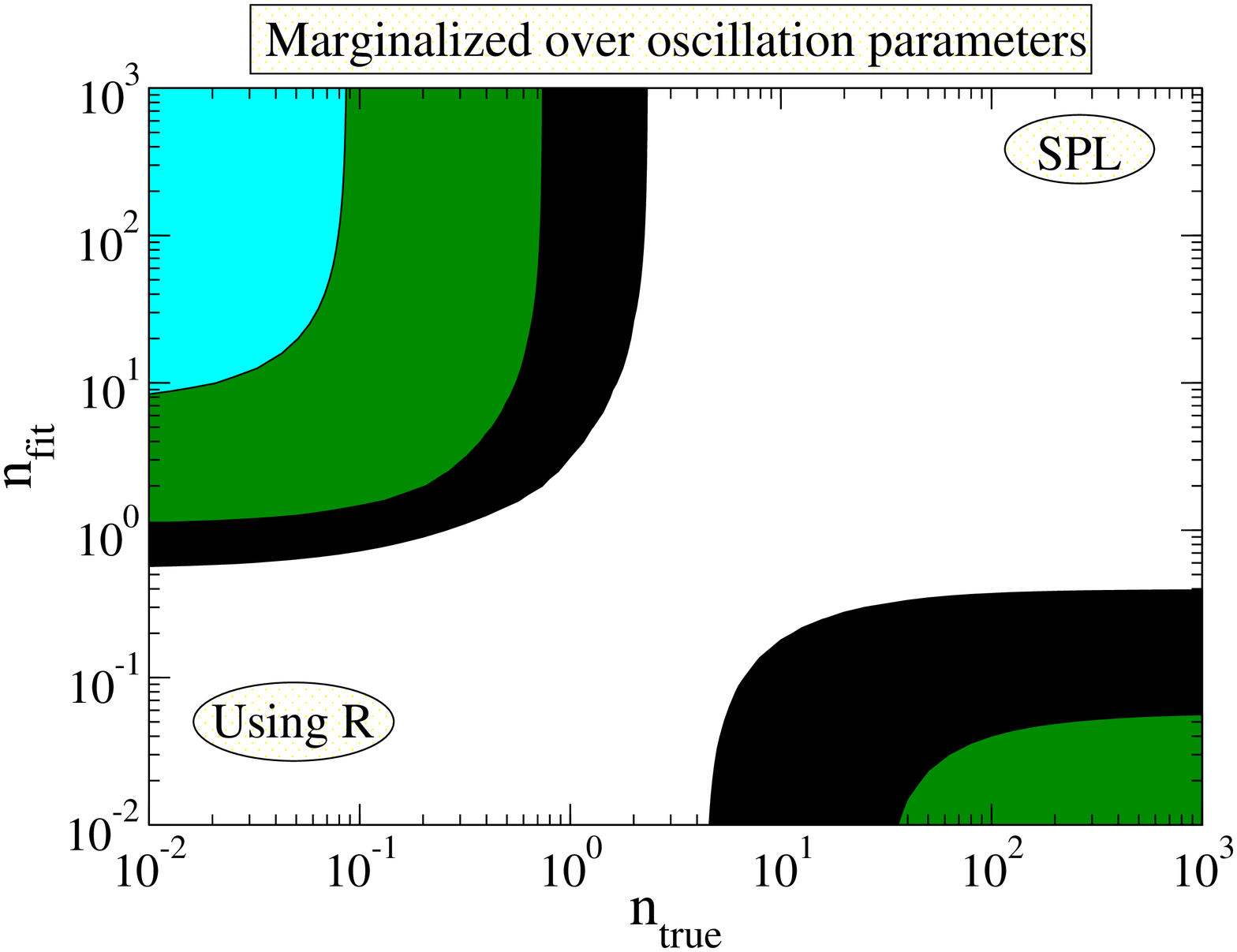}
\caption{\label{fig:ntruen}
For each $n_{true}$ on the $x$-axis, 
the shaded region shows the values of  
$n_{fit}$ which can be excluded at $2\sigma$. 
We assumed 
10\% (black/darkest), 20\% (green/dark) and 30\% (cyan/lightest) 
uncertainty on $\Delta R/R$. The four panels are for the four 
benchmark values of the oscillation parameters assumed in the 
data. 
}
\end{center}
\end{figure}

{ Now we allow the 
oscillation parameters to vary freely in the fit -- 
the only restriction being that they are not allowed to take 
values beyond their current $3\sigma$ limits.  
We find that, as expected from Fig.~\ref{fig:TR}, 
there is absolutely no $n$ sensitivity if $T$ is used.   
The reason, as discussed before, 
is the following: for TBM mixing parameters $T_{Data}=1/3$ for 
$n_{true}=2$. This 
value of $T$ can be reproduced by all values of $n$, as long 
as we are allowed to pick the most suitable set of 
oscillation parameters which lie within the current $3\sigma$ limit. 
Therefore, for this case the $T_{Theory}$ always exactly reproduces 
the $T_{Data}$ and thus $\chi^2=0$ for all $n_{fit}$. For $R$ 
the situation is slightly better. $R_{Data}=1$ for 
TBM mixing and $n_{true}=2$. And hence, as predicted from 
Fig.~\ref{fig:TR}, values of $n_{fit} < 1$ can still 
be disfavored. The reason being that for $n_{fit} < 1$, it 
is impossible to get the $R_{Theory}$ close to 1.
However, for $n_{fit} > 1$, one can always find 
a set of oscillation parameters within the still allowed region 
which gives the same $R_{Data}$.  
Fig.~\ref{fig:ntruefree} shows the result of the fit.}
\\

\begin{figure}
\begin{center}
\includegraphics[width=6cm,height=7cm,angle=270]{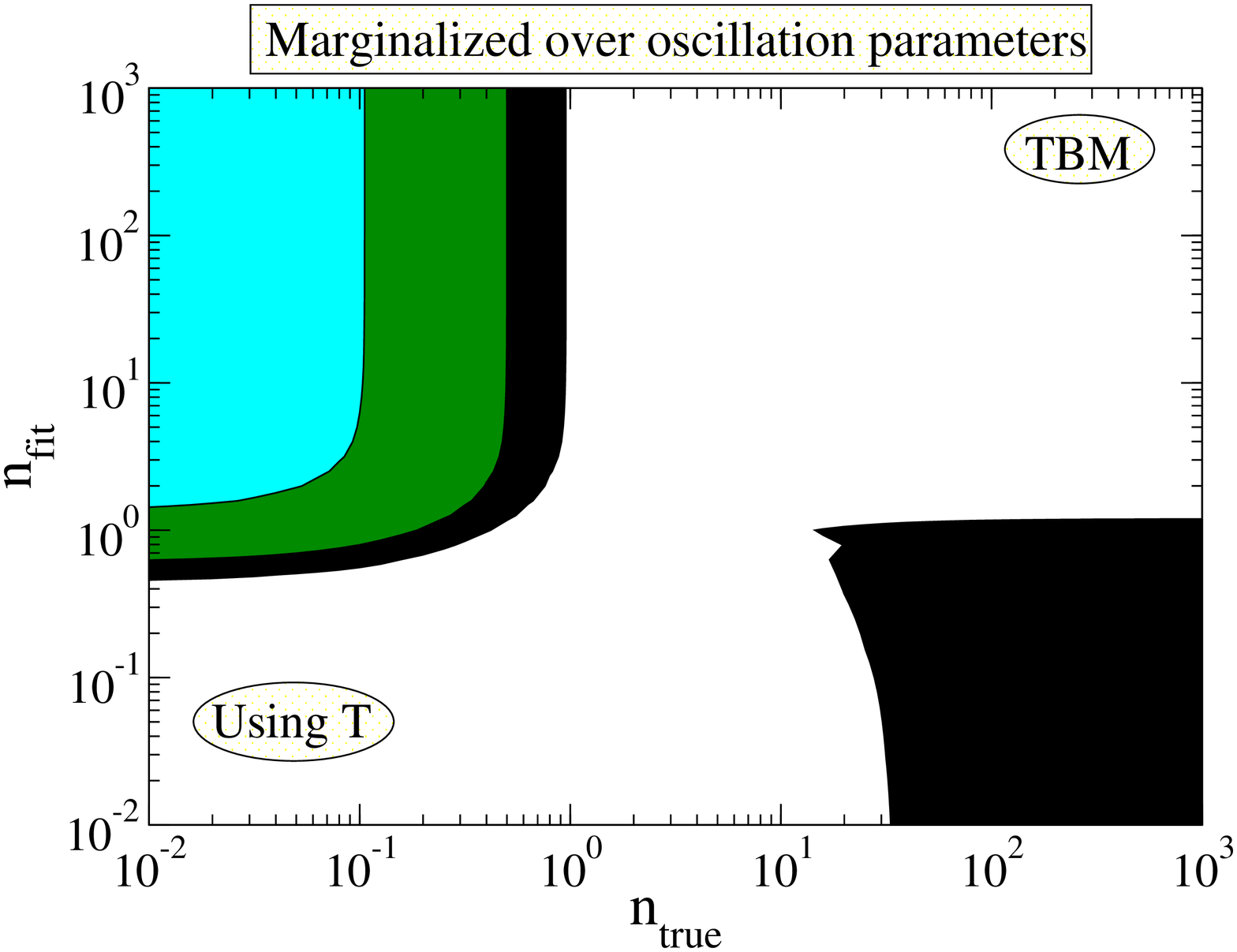}
\includegraphics[width=6cm,height=7cm,angle=270]{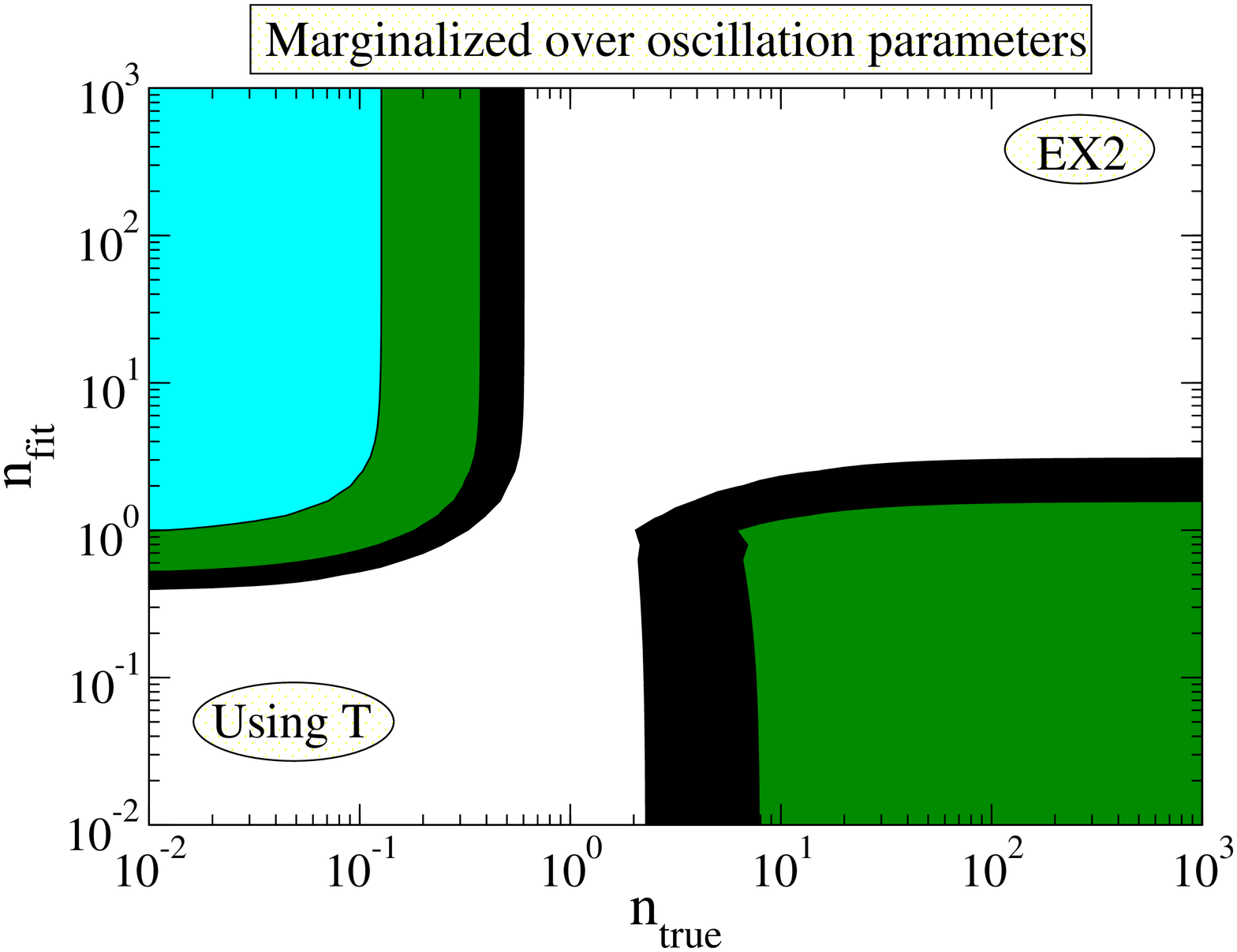}
\caption{\label{fig:ntruenT}
For each $n_{true}$ on the $x$-axis, 
the shaded region shows the values of  
$n_{fit}$ which can be excluded at $2\sigma$. 
We assumed 5\% (black/darkest), 10\% (green/dark) 
and 20\% (cyan/lightest) uncertainty on $\Delta T/T$. 
The two panels are for the benchmark values TBM and EX2 
of the oscillation parameters assumed in the 
data. 
Scenarios EX1 and SPL have no
sensitivity at all. 
}
\end{center}
\end{figure}

So far we have shown all $\chi^2$ sensitivity results assuming 
that we had $n_{true}=2$. We next show $2\sigma$ contour plots 
in the $n_{true}-n_{fit}$ plane in Figs.~\ref{fig:ntruen} and 
\ref{fig:ntruenT}. The way these figures are to be interpreted 
is the following: for each $n_{true}$ on the $x$-axis, 
the shaded region shows the values of 
$n_{fit}$ which can be excluded at $2\sigma$. 
Fig.~\ref{fig:ntruen}
shows the sensitivity on $n$ of the neutrino telescope using 
the more powerful flavor ratio $R$, while Fig.~\ref{fig:ntruenT} 
gives the corresponding reach by using the more easily 
measurable $T$. The darkest (black) regions are 
obtained assuming that we have 10\% (5\%) uncertainty in 
$R$ ($T$), the dark (green) regions are for an uncertainty 
of 20\% (10\%) in $R$ ($T$), while the lightest (cyan) regions 
are for uncertainty of 10\% (5\%) in $R$ ($T$).  
We show these regions in Fig. \ref{fig:ntruen} 
for the four benchmark sets 
of mixing parameter cases that we have considered in this paper. 
The upper left hand panel is for TBM mixing, the upper right hand 
panel is for the ``EX1'' oscillation scenario, the lower 
left hand panel is for the ``EX2'' oscillation case and the 
lower right hand for the special case (``SPL'') 
of trimaximal mixing in the 
second row. For $T$ we show the cases for ``TBM'' and ``EX2'' 
only in Fig.~\ref{fig:ntruenT}. 
The way we have generated these figures is the following: 
for each value of $n_{true}$ we generate the data for the particular 
oscillation parameter set. For instance for the upper left hand 
panel of Fig. \ref{fig:ntruen} 
the data is always generated for the TBM mixing parameters. 
This data is then fitted with any value of $n_{fit}$, 
while all oscillation parameters are allowed to vary 
within their $3\sigma$ allowed 
range. From Fig.~\ref{fig:ntruen} we 
find that the neutrino telescopes have reasonable 
sensitivity to the initial flux composition if $R$ can be 
measured with reasonable precision. The sensitivity 
depends sharply on the true value of the oscillation parameters 
as well as on the true value of $n$. We also note that irrespective 
of the true oscillation parameter scenario, there is a narrow 
band around $n_{true} \sim 1$, where we have no sensitivity to the 
initial flux composition. 
From Fig.~\ref{fig:ntruenT} we see 
that for TBM mixing and 
EX2, $T$ returns a sensitivity which is not bad. In particular, 
for TBM mixing, which is the currently favored scenario, 
one finds reasonable sensitivity to the flux composition 
if $n_{true}\ls 1$. On the other hand we have 
numerically checked that if the true neutrino 
mixing turns out to be compatible with the EX1 and SPL case, 
then it would be almost impossible to determine the 
UHE neutrino initial flavor composition using $T$ alone. 
In other words, there are no shaded regions for 
SPL and EX1 in the $n_{true}-n_{fit}$ space. This is why 
we do not show the panels for these cases in Fig.~\ref{fig:ntruenT}. 
\\

Of course it is expected that the current uncertainties on the 
oscillation parameters will decrease \cite{precise},
as better neutrino oscillation 
experiments are built and more data is accumulated. 
This would improve further the prospects of measuring $n$ at 
neutrino telescopes.

\section{\label{sec:concl}Conclusions and Summary}

Observation of neutrinos with very high energies
coming from astrophysical sources has been long overdue.
The observed UHE fluxes have a flavor composition on Earth 
which depends on both their initial composition at source 
as well as on neutrino flavor oscillations. 
Therefore, it should be possible to study both kinds of physics 
with these observations -- neutrino physics as well as 
physics concerning the UHE neutrino sources. 
Unfortunately, since there are uncertainties in both 
neutrino mixing parameters as well as knowledge on 
the type of sources of these UHE neutrinos, attempts on 
determining one are always plagued by uncertainties on 
the other. 

In this paper we attempted to probe the 
potential of the future neutrino telescopes in 
deciphering the flavor composition of the UHE neutrino 
flux at source. 
We did this by proposing a very simple and minimal 
parametrization of the flavor composition at source. We 
parameterized the 
initial flux composition of high energy neutrinos simply as 
\be
(\Phi_e^0 : \Phi_\mu^0 : \Phi_\tau^0) = (1 : n : 0) \, .
\ee
The parameter $n$ can take any value from $0$ to $\infty$. 
Specific values of $n$ describe the pure known sources. 
The 
UHE neutrino flux depends on a number of astrophysical factors and 
its exact flavor composition 
will depend on the process which generates 
this flux. Neutrinos coming from pion decay will have a 
flavor ratio of $(1:2:0)$. Likewise a muon-damped source will 
give a ratio of $(0:1:0)$, a neutron beam source will give $(1:0:0)$ 
and a charm source $(1:1:0)$. 
One expects impurities in these relations, i.e., they will not be
in the exact form given above.  
The astrophysical source producing the 
UHE neutrinos could have a 
combination of some or all such processes producing these neutrinos. 
Also, unless there are close-by astrophysical sources,  
what will be detected at the neutrino telescope is most likely 
going to be a diffuse flux coming from a combination of different 
astrophysical sources, and hence it is expected that this 
observed flux would come as 
a combination of different neutrino generating decay processes. 
Since the  
number of $\nu_\tau$ produced at the source is negligible, our 
one parameter description is the most economical and general 
way of probing the initial flavor composition of the UHE neutrinos. 

The observed flavor ratios depend on the neutrino mixing parameters 
in addition to the property of the source. 
We studied two of the most used observed flux 
ratios $T=\Phi_\mu/(\Phi_e + \Phi_\mu + \Phi_\tau)$ and 
$R=\Phi_e/\Phi_\tau$. 
We wrote down approximate analytic forms for these flux 
ratios in terms of $n$ and the mixing parameters. Using exact 
numerical results 
we showed the uncertainty in $T$ and $R$ coming from the 
current uncertainty on the mixing parameters. It was shown that the 
uncertainties due to mixing parameters in minimum around 
$n\simeq 1-2$. 
We also showed that for both $T$ and $R$ the uncertainty due to 
mixing for small $n$ was more than that for large $n$. We also 
pointed out a special case where $T = 1/3$ for all values of $n$. 
This scenario corresponds to the case where the second row of the 
neutrino mixing matrix has equal entries in all its three elements and
illustrates that the ratio $T$ alone may not suffice to probe $n$. 
\\

We next defined a simple $\chi^2$ function and expounded the 
potential of the neutrino telescopes in determining $n$ and 
hence the UHE neutrino flux composition at source. 
Since the forthcoming neutrino telescopes are yet to collect 
any data on ultra high energy neutrino 
fluxes, and since it is not yet known 
what kind of total uncertainty we would have on the observed 
flux ratios, 
we performed the $\chi^2$ analysis for four benchmark points in the 
mixing parameter space and by assuming different values of 
errors on $T$ and $R$. 
Those errors cover assumptions ranging from plausible to
optimistic, and allow to compare the prospects of 
neutrino telescopes. 
In particular, our analysis shows what kind of  
statistics and systematics are required from neutrino telescopes 
in order to achieve UHE source flavor sensitivity.  
We first studied the case by 
assuming a pure pionic source. It was seen that in this case, 
once the uncertainties 
due to oscillation parameters were taken into account, 
$T$ as measured in neutrino telescopes could 
give absolutely no information about the flux composition at source. 
The measured $R$ could still be used to exclude certain 
ranges of $n$ 
and hence certain types of sources. Finally, we performed a full 
scan of the $n$ space, where data was generated at every value of 
$n_{true}$ and the potential of the data to pick the right source 
$n$ was studied.  We presented $2\sigma$ contours 
in the $n_{true}-n_{fit}$ 
parameter space,  which give the $2\sigma$ initial 
flux composition sensitivity of the 
neutrino telescope, as a function of $n_{true}$. \\

In conclusion, neutrino telescopes will provide information on the 
flavor composition of the UHE neutrino fluxes on Earth. This can 
be used to study the initial flavor composition of these fluxes at 
their source. Reasonable sensitivity to deciphering the correct 
source flavor composition is expected from these experiments, 
despite the current uncertainties on the mixing parameters. With 
projected improvements in our understanding of the neutrino mixing 
angles, the source flavor sensitivity of the neutrino telescopes will 
improve.

\vspace{0.3cm}
\begin{center}
{\bf Acknowledgments}
\end{center}
We thank Sergio Palomares-Ruiz for helpful comments. 
W.R.~was supported by the ERC under the Starting Grant 
MANITOP and by the Deutsche Forschungsgemeinschaft 
in the Sonderforschungsbereich 
Transregio 27 ``Neutrinos and beyond -- Weakly interacting 
particles in Physics, Astrophysics and Cosmology''.  
S.C.~wishes to thank the Max--Planck--Institut f\"ur Kernphysik, 
Heidelberg, where part of this work was completed and 
acknowledges support from the Neutrino Project
under the XI Plan of Harish--Chandra Research Institute.

\pagestyle{empty}

\end{document}